\documentclass{aastex61}
\usepackage{epstopdf}
\def\gsim{\ \raise 3pt \hbox{$\rangle$} \kern -8.5pt \raise -2pt \hbox{$\sim$}\ }
\usepackage{graphicx,natbib,color,amsmath,subfigure}
\usepackage{ulem}
\newcommand{\blank}[1]{}
\usepackage{color}
\usepackage{ulem}

\usepackage{multirow}

\def\rhessi{{\textit{RHESSI}}}
\def\trace{{\textit{TRACE}}}
\def\kw{{Konus-\textit{Wind}}}
\def\mw{{microwave}}
\def\Mw{{Microwave}}

\begin{document}

\title{ ELECTRON ACCELERATION AND JET-FACILITATED  ESCAPE  IN AN M CLASS SOLAR FLARE ON 2002 AUGUST 19 }

\correspondingauthor{Lindsay Glesener}
\email{glesener@umn.edu}

\author{Lindsay Glesener}
\affil{School of Physics and Astronomy \\
University of Minnesota \\
116 Church St. SE \\
Minneapolis, MN 55455, USA}

\author{Gregory D. Fleishman}
\affiliation{Center for Solar-Terrestrial Research \\
New Jersey Institute of Technology \\
Newark, NJ 07102, USA}

\begin{abstract}

Sudden jets of collimated plasma arise from many locations on the Sun, including active regions. The magnetic field along which a jet emerges is
often  open to interplanetary space, offering a clear ``escape route'' for any flare-accelerated electrons, making jets lucrative targets for studying particle acceleration and the solar sources of transient heliospheric events. Bremsstrahlung hard X-rays (HXRs) could, in principle, trace the accelerated electrons that escape along the paths of the jets, but measurements of the escaping electron beams are customarily difficult due to the low densities of the corona.
In this work, we augment HXR observations with gyrosynchrotron emission observed in microwaves, as well as extreme ultraviolet (EUV)
emission and modeling to investigate flare-accelerated electrons in a coronal jet.  HXR and microwave data from \rhessi\, and OVSA, respectively, give complementary insight into electron spectra and locations, including the presence of accelerated electrons in the jet itself.  High-time-resolution HXR data from the \kw\, instrument  suggest electron acceleration timescales on the order of 1 second or shorter.  We model the energetic electron distributions in the GX Simulator framework using \textit{SoHO}/MDI, \rhessi,  \trace, and OVSA data as constraints.  The result is a modeled distribution, informed and constrained by measurements, of accelerated electrons as they escape the Sun. Combining  the detection of microwave gyrosynchrotron emission  from an open, rather than closed, magnetic configuration, with realistic 3D modeling constrained by  magnetograms, EUV, and X-ray emission, we obtain the most stringent constraints to date on the accelerated electrons  within a solar jet.
\end{abstract}


\section{Introduction}

At the Sun, transient events on magnetic field open to interplanetary space manifest themselves as jets, i.e. collimated, sudden ejections of chromospheric or coronal plasma.  These are commonly thought to arise by interchange reconnection between closed and open field, allowing plasma and accelerated electrons to escape the lower corona.
Jets are fundamental solar phenomena, occurring in all layers of the atmosphere and all regions (coronal holes, active regions, and the quiet Sun).  These events often offer a convenient magnetic configuration in which to study solar flare particle acceleration, particularly the acceleration of escaping energetic electrons.

  In the Shibata two-dimensional model of interchange reconnection, emerging flux reconnects with the overlying coronal field and causes the open field to switch footpoints \citep[e.g.][]{shibata1992}.  This occurrence can drive a hot (several MK) jet generated in the corona at the location of the upper reconnection outflow shock, and/or a relatively cooler jet of chromospheric temperature via sudden chromospheric evaporation  \citep{yokoyama1995, yokoyama1996}.  In the 3D model of \citet{pariat2015}, the jet arises in a fan-spine geometry, with reconnection occurring at a separatrix layer.  This model can generate straight, linear jets, in which the ejection is simple and rotationless, but these straight jets quickly devolve into more complicated, untwisting, helical jets.  The untwisting arises as the newly open field, no longer constrained by two photospheric footpoints, is able to shed its twist, and the propagation of this twist upward can help to drive the jet.  \citet{pariat2015} surmise that these two types of jets, straight and helical, might account for the dichotomy identified by \citet{moore2010}, which those authors called ``standard'' and ``blowout'' jets.  \citet{sterling2015, sterling2016} have proposed that jets are miniature filament eruptions, akin to a small-scale CME, and \citet{wyper2017} have performed 3D simulations suggesting that CMEs and jets result from the same processes.

Jets are often detected at soft X-ray (SXR) wavelengths \citep{shimojo1996, shimojo2000}, with high occurrence rates in the polar regions and coronal holes \citep{sako2012, savcheva2007}, presumably because of the prevalence of open field.
Jets in active regions may be accompanied by flares \citep[e.g.][]{kundu1999} and can accelerate electrons to energies at least up to hundreds of keV.  These accelerated electrons travel along the path of the jet, escaping the low corona, as indicated by the strong correlation between jets and Type III radio bursts \citep[e.g.][]{kundu1995, raulin1996}, with Type III sources appearing along the jet path in order of decreasing frequency (i.e. decreasing density) with height.  Jets are sometimes associated with impulsive, electron/He$^3$-rich solar energetic particle (SEP) events \citep{wang2006, nitta2015}, emphasizing that jets accelerate particles on field open to interplanetary space.

Gyrosynchrotron microwaves and bremsstrahlung hard X-rays (HXRs) are complementary tools for studying flare-accelerated electron distributions, as they can both quantitatively measure those distributions.  In principle, both should be useful in studying electrons accelerated in jets, but in practice neither is typically observed from coronal jet tracks due to observational difficulties.  HXR observations are dominated by footpoint emission in the (dense) chromosphere from downward-directed beams; emission from escaping electron beams in the low-density corona is typically too faint to be observed \citep{saint-hilaire2009}.  Gyrosynchrotron observation requires a ground-based observatory measuring at the right time and right frequency range for the magnetic field specific to each event, and so is also usually not observed from jets.  Type III radio emission is an excellent qualitative marker of accelerated electrons and their paths \citep[e.g.][]{reid2014, chen2013}, but due to the nonlinear processes in its generation cannot be used to quantitatively measure the emitting electron distributions.

The few jets studied in HXRs shed insight on the physical mechanism. \citet{krucker2011} studied jet HXR footpoints and found the configuration to be consistent with interchange reconnection; most of the 16 events, identified via prompt in-situ electron detections, exhibited three footpoints, as opposed to typical two-ribbon flares.  All HXR sources observed in that study were footpoint sources due to downward-directed electron beams impacting the chromosphere; any coronal sources in the jet were too faint to be observed.  \citet{bain2009} and \citet{glesener2012} did find coronal HXRs in jets, emphasizing the accelerated electrons' access to open field.  These rare observations were made possible due to an extraordinarily large electron flux in the former case and, in the latter, the fact that the flare footpoints were occulted by the solar limb.  The partly occulted observation revealed a double coronal source early in the flare, suggesting a reconnection site between the sources, followed by a relatively intense, extended HXR source cospatial and cotemporal with the emerging jet.  These HXRs imply accelerated electrons near the base of the jet, lower in altitude than the reconnection region, but it is not clear how electrons might access this site from the reconnection region.
  In that event, HXR emission was not strong enough for detailed spectroscopy to precisely determine the parameters of the emitting electron distribution.

\begin{figure}[tb]
\centering
\quad\includegraphics[width=0.6\columnwidth,trim=0 3cm 0 0,clip]{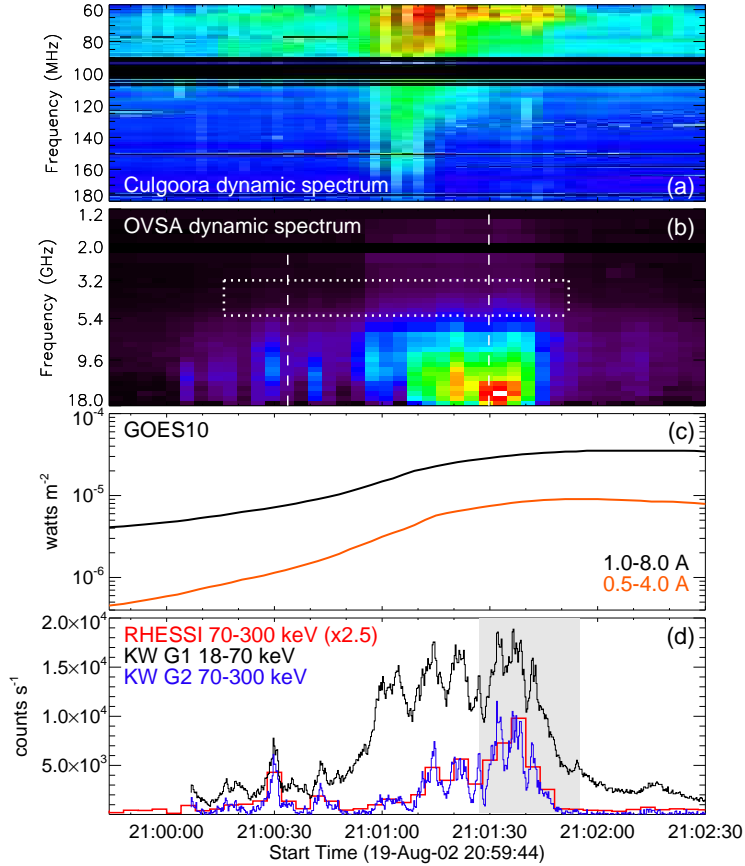}
\caption{\label{f_20020819_overview} Overview of the 2002 August 19 flare. (a) Dynamic spectra from the Culgoora Radio Observatory, showing Type III radio bursts.  Although Type III emission is prevalent in this jet, the most intense plasma emission is not concurrent with the most intense broadband emission.  (b) Broadband emission observed in OVSA microwaves, showing strong and quickly changing spectral variability. The dotted rectangle indicates the range of  frequencies and times used to obtain the \mw\ image shown in Figures~\ref{f_Jet_2002_08_19_images} and \ref{f_Jet_2002_08_19_OVSA_decon}. The vertical dashed lines indicate the times for the instantaneous spectra
 used in our 3D modeling; (c) SXR light curves in two \textit{GOES} channels.
(d) HXR count time profiles from \kw\, and \rhessi. The \rhessi\, curve includes its 8 segmented detectors (out of 9) and the counts have been scaled for comparison with \kw. The shaded area indicates the time interval used for \rhessi\ imaging shown in Figure~\ref{f_Jet_2002_08_19_images}.
}
\end{figure}

In this work we continue the progression of knowledge on accelerated electrons escaping the Sun in jets.  While context measurements from Type III observations and in-situ studies have long established the existence of these escaping electrons, it was not until the HXR studies of \citet{glesener2012} and \citet{bain2009} that energetic estimates of the escaping populations could begin to be performed.  These limited previous observations rarely allowed detailed spectroscopy due to limited HXR imaging dynamic range, and did not have cotemporal spectrally and spatially resolved microwave observations.  In this work, we combine imaging and spectral HXR observation by the \textit{Reuven Ramaty High Energy Solar Spectroscopic Imager} (\rhessi)  with high cadence spectroscopy from \kw, \mw\ spectral and imaging data from the Owens Valley Solar Array (OVSA), and extreme ultraviolet (EUV) data from the \textit{Transition Region and Coronal Explorer} (\trace).  We do not know of any previous flare-related jet that has been studied using this set of observational tools.  The observations are augmented by 3D modeling utilizing photospheric line-of-site measurements of the magnetic field and linear force-free reconstruction of the coronal magnetic environment for one of the 16 jet events reported by  \citet{krucker2011}.  With these tools we investigate accelerated electron distributions on open field, the relation of these accelerated electrons to the flare and the jet, and fast spectral changes in the HXRs and microwaves emitted by these particles.  The GX Simulator framework \citep{Nita_etal_2015, 2018ApJ...853...66N} is used to develop 3D models of the flare and jet sources, which include open and closed magnetic flux tubes and distributions of thermal and nonthermal electrons within the sources. These models are validated via comparison with X-ray, EUV, and \mw\ data.  The work demonstrates the use of HXRs and microwaves together to \textit{quantitatively} constrain accelerated electron distributions in a jet.

\section{Observations}

\begin{figure*}\centering
\quad\includegraphics[width=0.95\textwidth,clip]{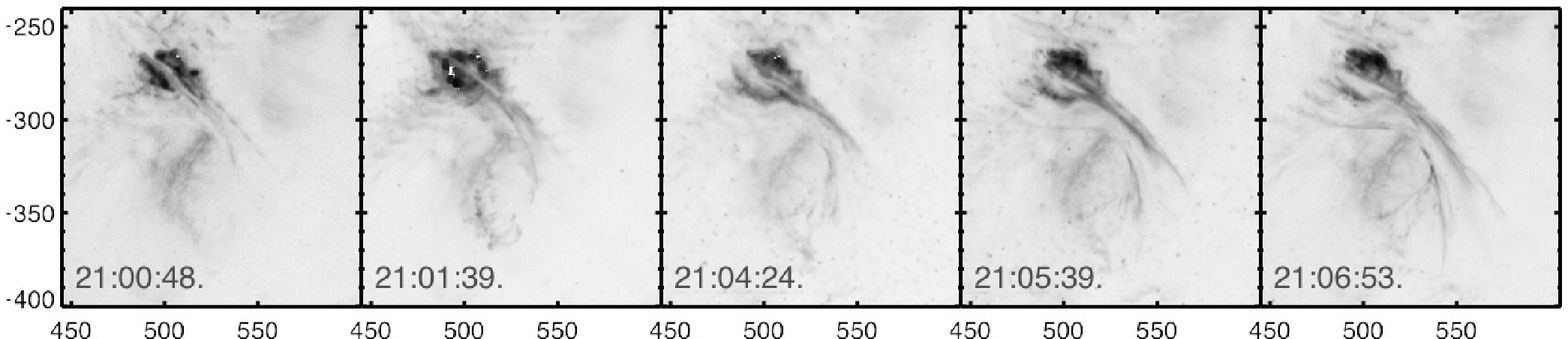}
\includegraphics[width=\textwidth,clip,angle=0]{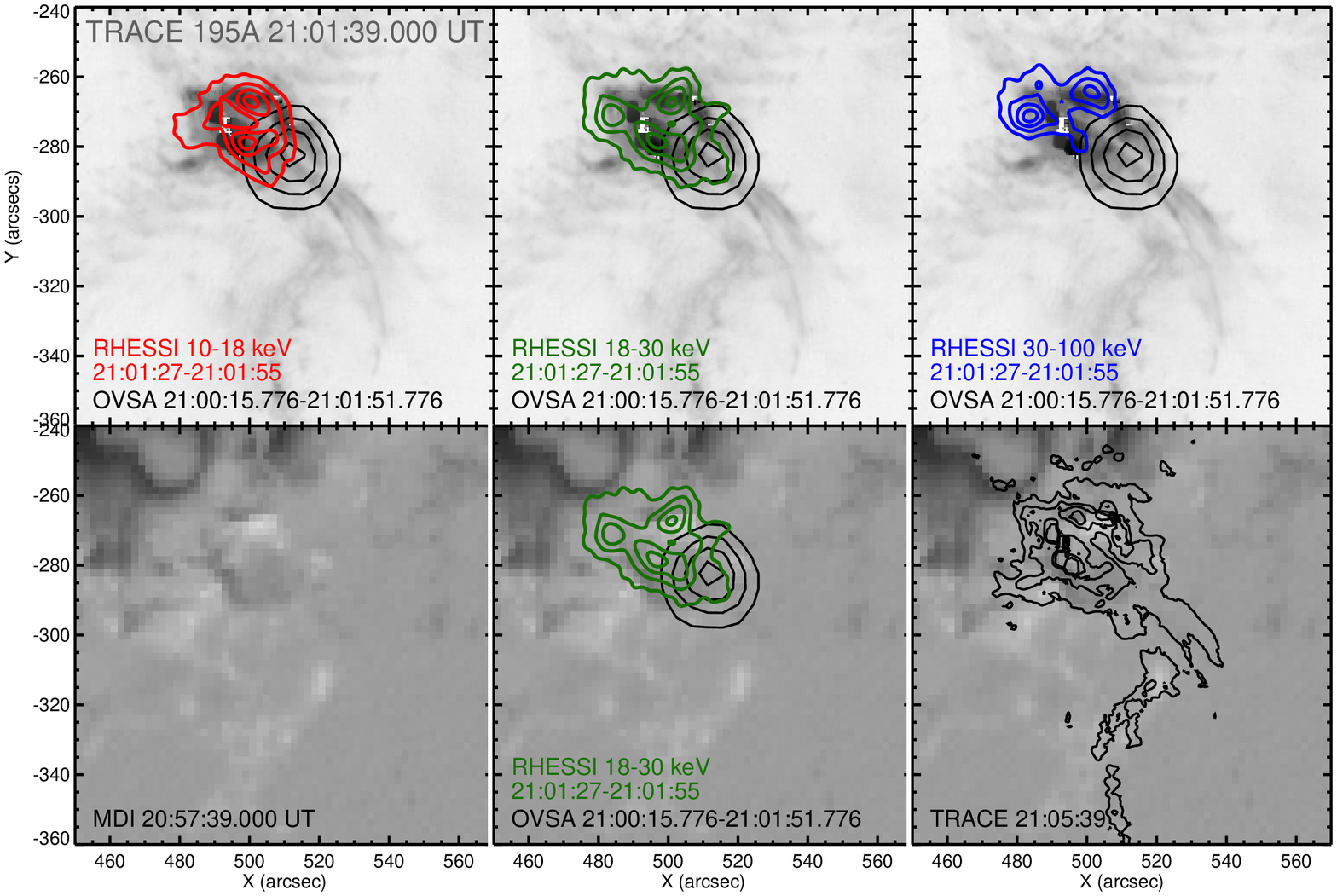}
\caption{\label{f_Jet_2002_08_19_images} Top panels: \trace\, images at 195\AA\, as the jet evolves.  (An animated form of these panels is available in the supplementary materials).  Middle row: \rhessi\, and OVSA contours overlaid on a \trace\, 195\AA\, image of the emerging jet.  All \rhessi\ and OVSA contour levels are 30, 50, 70, and 90\% of their respective maxima, and \rhessi\, images were produced using the CLEAN algorithm.  At low energies, HXR emission is thermal and likely emanates from flaring coronal loops, while the highest energies are nonthermal and probably denote footpoints at the base of the flare/jet.  In the intermediate range (18-30 keV), some HXR emission is elongated along the jet.  Bottom row:  \rhessi\ and \trace\ emission overlaid on MDI magnetograms.
 }
\end{figure*}

\subsection{EUV data from \trace}
\label{S_EUV_data}

The \trace\ spacecraft was an EUV imager that operated from 1998 to 2010 with an 8.5 arcminute field of view and a spatial resolution of 1 arcsecond.  \trace\, measured flux in three EUV and several UV wavelengths sensitive to selected temperatures from 6000 K to 10 MK \citep{trace}.  On 2002 August 19, \trace\, observed a solar jet associated with a flare from active region 10069 (see \textit{GOES} curve in Figure \ref{f_20020819_overview}), with coverage in the 195~\AA\ filter at approximately 23 second cadence during the impulsive part of the event.  This passband is sensitive to Fe XII and Fe XXIV lines with peak temperature sensitivity at log[T(MK)]=6.2 and 7.2, respectively \citep{landi2013}.  The top set of panels in Figure~\ref{f_Jet_2002_08_19_images} shows \trace\, snapshots of the jet.  Some panels evidence saturation and diffraction during the bright flare, which had a \textit{GOES} class of M3.1.  EUV jet emission continued for several minutes after the initial bright phase.

\trace\, pointing knowledge is not precise and could be incorrect by a few arcsec \citep{trace}.  For the 2002 August 19 event, there are no context observations at a similar time and wavelength that can be used for absolute calibration of this pointing.  Instead, we coaligned quiescent Fe XII plage features observed by \trace\, to \textit{SOHO}/MDI magnetic data.  The primary feature utilized can be seen in Figure \ref{f_Jet_2002_08_19_images} extending southeast from $\sim$[520, -310] arcseconds.  All \trace\ images shown in this paper include this alignment correction.

\subsection{Hard X-ray data from \rhessi\ and \kw}
\label{S_Xray_data}

The flare/jet event was observed by the {\it RHESSI} spacecraft, which provides high-resolution X-ray spectra and full-disk images of the Sun from 3 keV to 17 MeV \citep{lin2002}.  \rhessi\, utilizes high-purity germanium detectors and rotation modulation collimation, an indirect, Fourier-based imaging system \citep{hurford2002}. \rhessi\, emission comes from two types of populations: hot thermal ($\gtrsim$10 MK) plasma and accelerated electrons.  The brightest nonthermal HXR sources customarily occur at flare footpoints; due to limited imaging dynamic range, \rhessi\, only occasionally observes fainter nonthermal sources in the corona.

\begin{figure*}[b]\centering
\includegraphics[width=\textwidth]{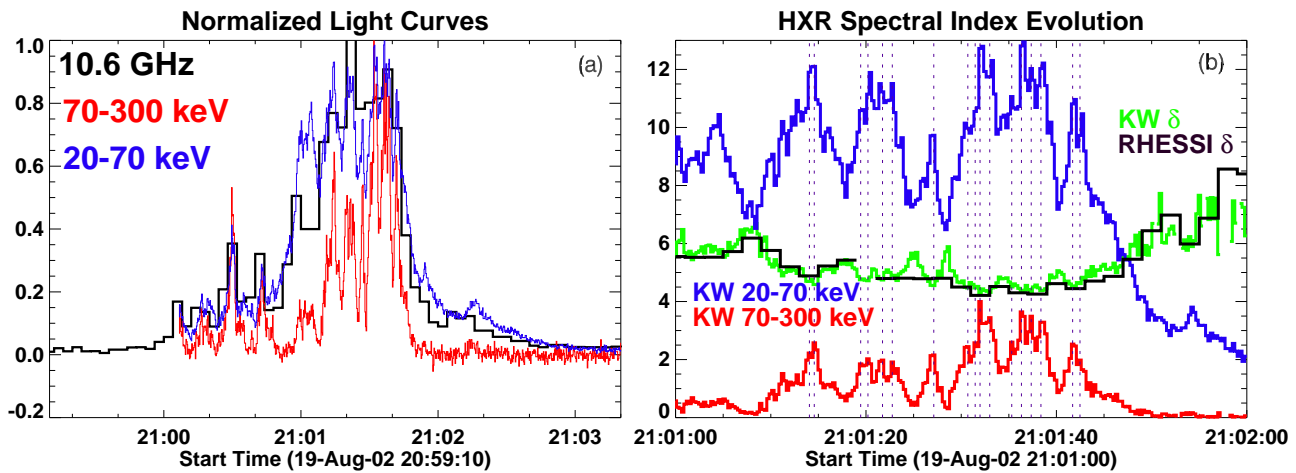}
\caption{\label{f_light_curves_OVSA_KW} HXR spectral evolution. (a) \kw\ HXR light curves obtained with 256~ms time resolution in wide energy channels G1 ($\sim$20--70~keV) and G2 ($\sim$70--300~keV). A 10.6 GHz OVSA light curve (shifted in time by 2.224~s to correct for OVSA clock error) with 4~s time resolution  is shown for comparison.  Sub-second time variability of the HXR emission is apparent. (b) (Green) evolution of the effective spectral index defined using the the \kw\ hardness ratio as explained in \citet{Fl_etal_2016coldFl}. This spectral index displays time variability similar to that of the HXR light curves (blue, red), with the spectrum hardening at most HXR peaks. On average, the effective spectral index agrees well with that determined from the \rhessi\ fit (black) in 2-second time bins.
}
\end{figure*}

\begin{figure*}\centering
\includegraphics[width=0.8\textwidth]{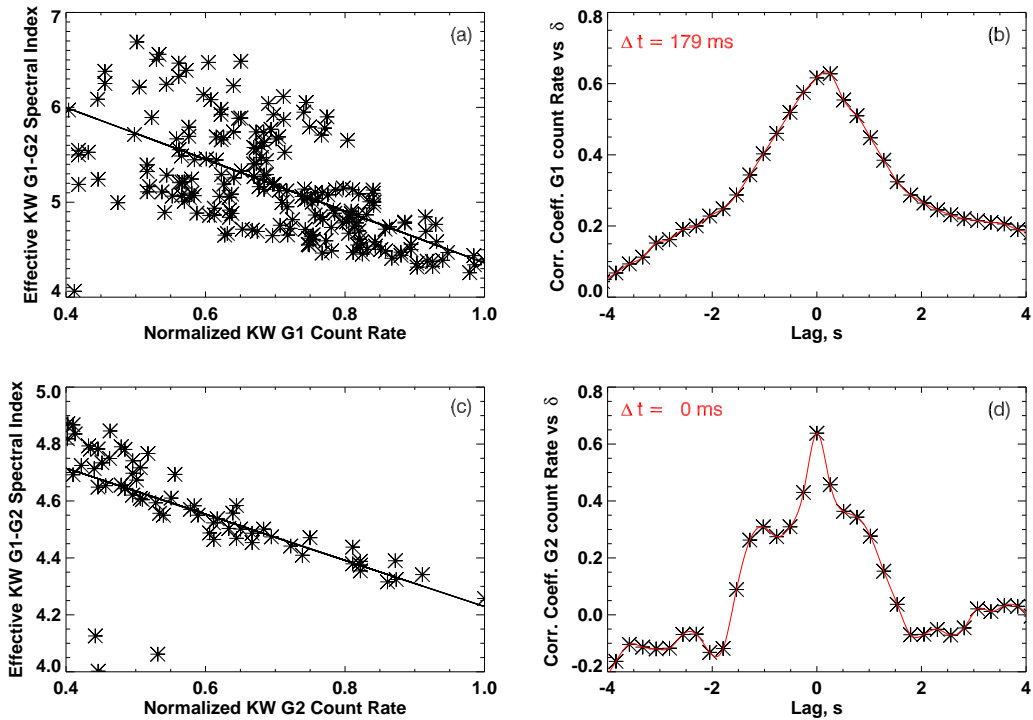}
\caption{\label{f_cross_correlation} Comparison of HXR flux and spectral index. Panel (a) shows the correlation between the \kw\ spectral index and the flux in channel G1 (20--70~keV), making evident the spectral hardening at times of high flux.  Panel (b) plots the cross-correlation between the two quantities.  Panels (c) and (d) show the same for channel G2 (70--300 keV). The spectral hardening is delayed relative to the G1 intensity by roughly 0.2~s, while there is no delay relative to G2 intensity.  }
\end{figure*}

The middle row of panels of Figure \ref{f_Jet_2002_08_19_images} shows \rhessi\, images in three energy ranges for the 2002 August 19 flare overlaid on a \trace\, 195~\AA\ image of the emerging jet.  Images were produced using the CLEAN method with subcollimators 1, 3, 4, 5, 6, and 7 and a clean beam width factor of 0.9, integrated over 28 seconds.  This set of subcollimators is chosen to elucidate fine structure in the flare and jet.  All three \rhessi\, images show contour levels at 30, 50, 70, and 90\% of their respective maxima.  At low energies (10--18 keV), HXR emission is dominated by the thermal flare, while the highest energies (30--100 keV) are nonthermal and most likely trace out footpoints at the base of the flare/jet. In the intermediate range (18-30 keV), HXR emission is elongated along the jet, as indicated by the 30\% green contour.  (No imaging was attempted below 10 keV because \rhessi's thickest attenuator was inserted.)  The \rhessi\, sources at the flare and the jet are not isolated enough to perform imaging spectroscopy to separate the sources.  X-ray power-law spectral fits to the spatially integrated emission were performed using the \rhessi\, OSPEX software and are used as upper limits on the emission from any sub-component; see Section \ref{S_modeling}.  Integrated spectra (not shown) indicate that 10--18 keV and 30--100 keV emissions are part of the thermal and nonthermal spectral components, respectively, but the spectral shape of the 18--30 keV emission cannot be determined.

While \rhessi\, provides detailed images and spectra throughout the evolution of the event, its time resolution is limited by its rotation modulation.  Straightforward time profiles (i.e. without attempting demodulation) can be produced on a cadence $\ge$ 2 seconds (half a spacecraft rotation).    For high-cadence spectroscopy (though not imaging), we turn to HXR data from the Konus instrument aboard the \textit{Wind} spacecraft \citep[\kw, in operation since 1994;][]{Aptekar1995, Palshin_etal_2014}.  \kw\, uses NaI(Tl) crystals to measure photons from astrophysical sources from $\sim$10 keV to 15 MeV.  \kw\, lightcurves in two high-energy bands are shown in Figure \ref{f_20020819_overview} and Figure \ref{f_light_curves_OVSA_KW}.  The instrument records energy spectra independently from the light curves \citep{Aptekar1995} using an adaptive spectrum accumulation duration based on the signal intensity; thus, the spectra are taken over a number of uneven time intervals. However, as has been demonstrated by \citet[][Eq.~4]{Fl_etal_2016coldFl}, the electron spectral index can be accurately recovered from the hardness ratio determined using two wide channels only, G1 and G2 (20--70 and 70--300 keV, respectively).
This ratio is available with a sub-second cadence of 16---256~ms.

\begin{figure}[htb]
\begin{center}
	\includegraphics[width=0.9\linewidth]{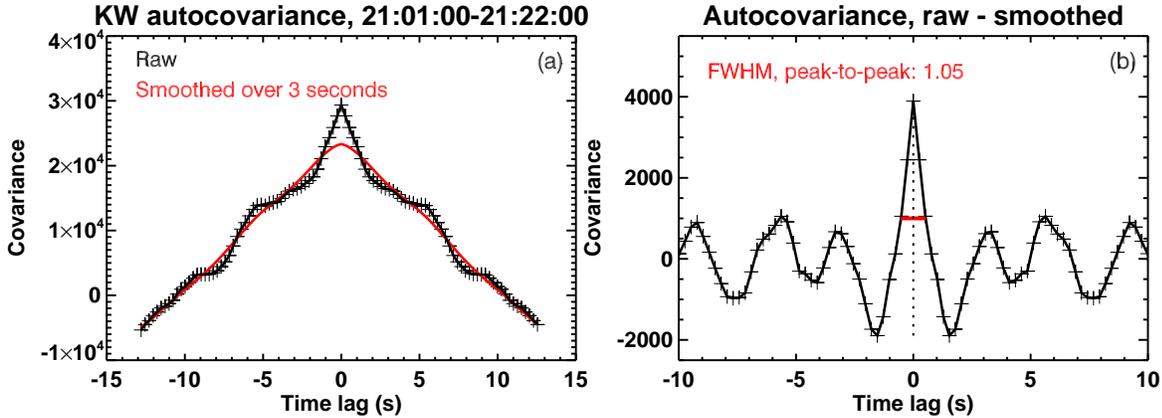}
\caption{ Autocovariance of the \kw\, 70--300 keV lightcurve for the 1-minute impulsive phase shown in Figure \ref{f_light_curves_OVSA_KW}, panel (b).  The average pulse duration is given by the width of the autocovariance peak.  (Left) Autocovariance for the raw \kw\, data and for data that have been smoothed over 3 seconds (to remove the effect of the fast time variation).  (Right) Autocovariance of the difference of the two time profiles (raw minus smoothed) isolates the contribution of the quickly-varying component.  A characteristic pulse duration is measured from the FWHM of this curve, which is 1.05 seconds, though note that individual bursts can be longer or shorter than this average duration.  The short pulses, which are too fast for \rhessi\, to resolve, could indicate electron acceleration timescales. }
\label{fig:autocorrelation}
\end{center}
\end{figure}

The time history of the \kw\ power-law index $\delta$ of the electron distribution is shown in the second panel of Figure \ref{f_light_curves_OVSA_KW}, with the \rhessi\, derived power-law index calculated at a two-second cadence overplotted.
Fast time variations down to subsecond timescales are apparent in the \kw\, intensity and spectral index.  Most notably, each HXR peak corresponds to a decrease in $\delta$, i.e. a momentary spectral hardening.  The relationship between HXR flux peaks and spectral hardening is explored more quantitatively in Figure \ref{f_cross_correlation}, where clear correlations are evident between the spectral index and the flux measured in \kw\, channels G1 and G2.  Figure \ref{f_cross_correlation} also shows a lag-correlation analysis, suggesting that the spectral hardness is delayed by roughly 0.2~s relative to the intensity peaks in G1 (20--70~keV), while the data are consistent with zero time lag between the intensity peaks in G2 (70--300 keV) and spectral hardening. This implies that a duration of $\sim 0.2$~s is needed for the electrons with energy $\gtrsim 20$~keV to be accelerated up to $\gtrsim 100$~keV.

Figure \ref{fig:autocorrelation} examines the timescales of the fast HXR fluctuations.  A typical pulse duration can be measured as the width of the autocovariance of the time profile.  The lefthand panel shows the autocovariance of the \kw\, 70--300~keV lightcurve for the 1-minute impulsive phase shown in Figure \ref{f_light_curves_OVSA_KW}, panel (b), as well as that for the same profile smoothed over 3 seconds (effectively a low-pass filter that removes the fast time variation).  In the righthand panel, an autocovariance for the difference of the raw and smoothed time profiles isolates the contribution of the quickly-varying component.  The FWHM of this curve (with height measured from peak to valley) is 1.05 seconds.  This is the average pulse duration; note that visual inspection of the lightcurve in Figure \ref{f_light_curves_OVSA_KW}, panel (b) shows that some peaks are subsecond.  At longer time scales, the autocovariance deviates from that of the smoothed curve at widths of 11.3 and 18.4 seconds, with smaller deviations at widths of 6.4 and 24.6 seconds.  The short pulses are too fast for \rhessi\, to resolve using traditional \rhessi\, 4- or 2-second time binning, making this a rare observation enabled by the use of \kw\, data.

\subsection{\Mw\ data from OVSA}
\label{S_radio_data}

\Mw\ data from OVSA are highly complementary to the \rhessi\ X-ray data  in that they offer a second method by which to measure flare-accelerated electrons that produce gyrosynchrotron emission. Before decommissioning in 2010, OVSA \citep{ovsa_1984, Gary_Hurford_1994} provided the total power data in the form of dynamic spectra \citep{Nita_etal_2004}, from which instantaneous spectra or frequency-specific light curves could be derived and analyzed, and interferometric data for imaging.
Recently, \mw\ imaging with OVSA has become more easily accessible\footnote{See detailed OVSA imaging manual at \url{http://ovsa.njit.edu/legacy/}.} with a newly-calibrated OVSA legacy data base\footnote{Currently available at \url{http://ovsa.njit.edu/data/archive/calibration_files/} from Jan 2000 to Aug 2003.} and updated OVSA imaging software \citep{2014AAS...22421845N}.

For our jet-associated microwave burst we employ both spectral and imaging OVSA capabilities, as briefly outlined in \citet[][Section~4.2]{Fl_etal_2016narrow} and described in the OVSA imaging manual in more detail.
 Using the SSW \verb"ovsa_explorer"  widget, we identified the burst time from a calibrated solar observation data set, subtracted the pre-burst background, and fit sequentially all instantaneous spectra using a built-in fit function.  For this fit we assume the broadband emission has a spectral shape consistent with a generic function that has the form \citep{1989SoPh..120..351S,Nita_etal_2004}

\begin{figure*}\centering
\includegraphics[width=0.35\textwidth, trim=0 5.6cm 8.2cm 0, clip=true]{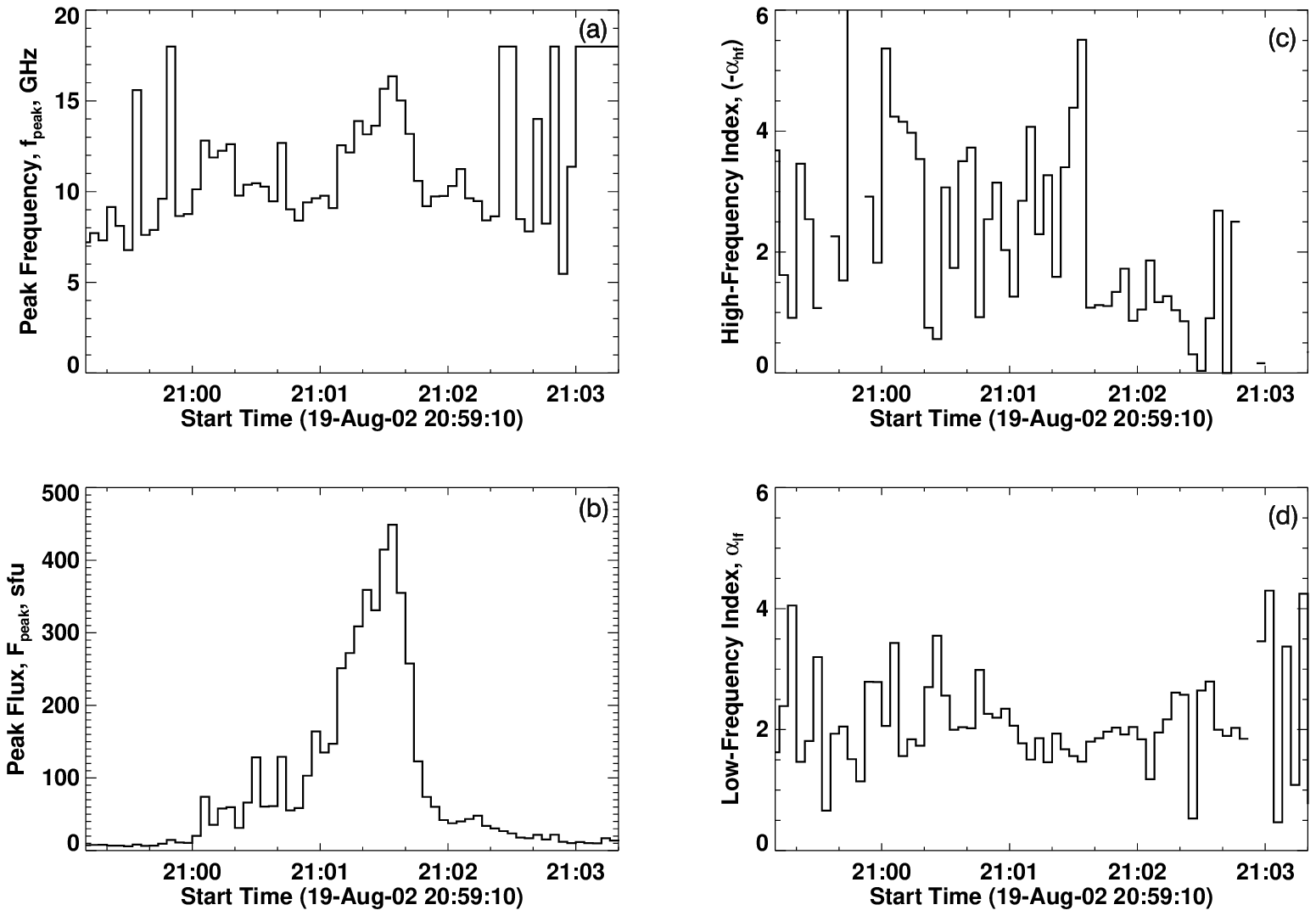}
\includegraphics[width=0.35\textwidth, trim=0 0cm 8.2cm 5.7cm, clip=true]{f6.eps}\\
\includegraphics[width=0.35\textwidth, trim=8.2cm 5.6cm 0cm 0, clip=true]{f6.eps}
\includegraphics[width=0.35\textwidth, trim=8.2cm 0cm 0 5.7cm, clip=true]{f6.eps}
\caption{\label{f_spec_evol_OVSA_KW}  Evolution of microwave spectral fit parameters, including (a) peak frequency, (b) peak flux, (c) high-frequency spectral index, and (d) low-frequency spectral index. Fast temporal variability is apparent.   }
\end{figure*}

\begin{figure}\centering
\includegraphics[width=0.45\textwidth]{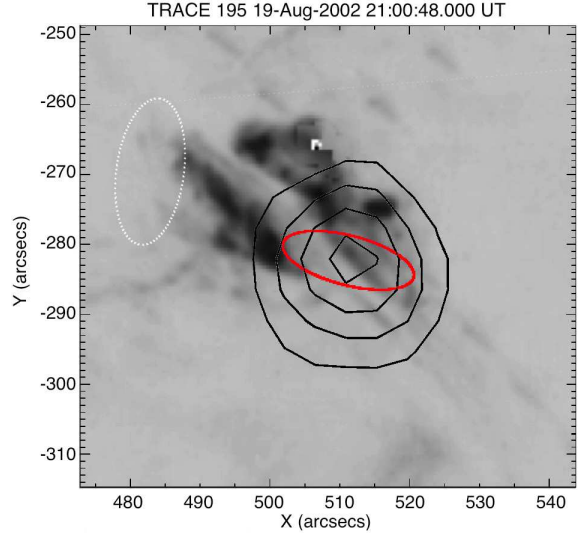}
\caption{\label{f_Jet_2002_08_19_OVSA_decon} { OVSA image synthesized over frequencies 3.2--5.4~GHz and 96 seconds at the burst peak, overlaid on the \trace\, jet.  Black contours show the OVSA image produced via the CLEAN$+$SELFCAL method, and the dotted white oval shows the CLEAN beam size (FWHM).  Deconvolution of the CLEAN beam from the image produces the elliptical source shown in red, which is elongated roughly in the direction of the jet, suggesting that the microwave emission emanates from the jet itself. 
}
 }
\end{figure}

\begin{equation}
\label{Eq_mw_fit}
S=e^{A} f^{\alpha}\left[1-e^{-e^{B} f^{-\beta}}\right],
\end{equation}
where $f$ is the frequency in GHz, and $A$, $B$, $\alpha$, and $\beta$ are the free fitting parameters that yield the physical parameters of interest. For example, the low-frequency spectral index is $\alpha_{\rm lf} \equiv \alpha$, while the high-frequency spectral index  is $\alpha_{\rm hf} = \alpha-\beta$. The fit results shown in Figure~\ref{f_spec_evol_OVSA_KW} reveal strong time variability in the spectral fit parameters---the spectral peak frequency, $f_{\rm peak}$, and the high- and low- frequency spectral indices, $\alpha_{\rm hf}$ and $\alpha_{\rm lf}$. This spectral parameter variability closely follows the \mw\ light curve variability seen in Fig.~\ref{f_light_curves_OVSA_KW}.  We note that this simple spectral fitting has assumed a single component to the spectrum.  Since this does not generally need to be the case, the 3D modeling described in Section \ref{S_modeling} will include multiple components and spatially varying magnetic fields.

The OVSA images are produced from the \verb"uv"-files using the IDL widget-based application \verb"wimagr."
 This application can generate images using CLEAN or CLEAN+SelfCal methods with various combinations of time and frequency synthesis, import other context images for comparison, and save the results in various formats.
   Most recently, the capability to analytically deconvolve a gaussian source model, which was employed in \citet{Fl_etal_2015} following \citet{1970AuJPh..23..113W}, has been added. Figure~\ref{f_Jet_2002_08_19_OVSA_decon} displays
in black contours an OVSA image synthesized over time  (96~s at the burst peak) and frequency (3.2--5.4~GHz)  obtained using the CLEAN+SELFCAL restoration method; it is overlaid on a \trace\,195~\AA\, image.  These time and frequency boundaries are also shown in the OVSA time profile in Panel (b) of Figure \ref{f_20020819_overview}. In addition, Figure~\ref{f_Jet_2002_08_19_OVSA_decon} shows a model gaussian deconvolved source (red) obtained using the built-in analytical deconvolution of a gaussian source and a gaussian beam proposed by \citet{1970AuJPh..23..113W}, where the effect of the finite and anisotropic beam of the array (dotted oval) has been removed.  To produce an OVSA image, high flux and precise, accurate phase determination are required.  Phase error increases with frequency, and we were not able to produce reasonable images above 5 GHz.  At low frequencies (e.g. 1--2 GHz), where the jet is more dominant, there is not sufficient flux to produce OVSA images.

The OVSA phase calibration employs observations of cosmic sources; thus, normally, the accuracy of source location is only limited by the phase fluctuations of the cosmic source (``calibrator'') measurement---typically, a few arcsec. No additional coalignment was applied to plot the images in Figure~\ref{f_Jet_2002_08_19_OVSA_decon}.
Comparison of the magnetic topology highlighted by the \trace\, image with the deconvolved \mw\ image suggests that the imaged \mw\ emission is elongated along the jet, rather than from a closed magnetic flux tube associated with the X-ray sources.
Indirect support for this possible role of the jet comes from the relative lack of signatures of trapping, which would broaden the temporal peaks in the radio emission time profile.

\section{3D modeling}
\label{S_modeling}

To facilitate solar 3D modeling and comparison with data the NJIT group developed a powerful tool, GX Simulator (Gyrosynchrotron/X-ray Simulator), which is now in SolarSoft (Nita et al. 2015, 2018).
 The tool allows the user to create a 3D model from an imported photospheric magnetic base map.  In GX Simulator various aspects of the model can be displayed, manipulated, or altered, such as magnetic field components, field lines, flux tubes, thermal and nonthermal density, and nonthermal energy parameters (e.g. power-law index, pitch angle, high/low energy cutoffs). We use the tool to build 3D models using realistic magnetic configurations obtained from force-free extrapolations of photospheric magnetograms from the Michelson Doppler Imager (MDI; see Figure \ref{f_Jet_2002_08_19_images} for the magnetogram in comparison with the flare/jet).
 We then compute emission in EUV, HXRs, and radio by numerically solving the corresponding radiative transfer equations.  We thus synthesize, from a single 3D model, multi-wavelength images and spectra and compare them with all available observed data to validate the model.

\begin{figure*}\centering
\includegraphics[width=0.75\textwidth,clip]{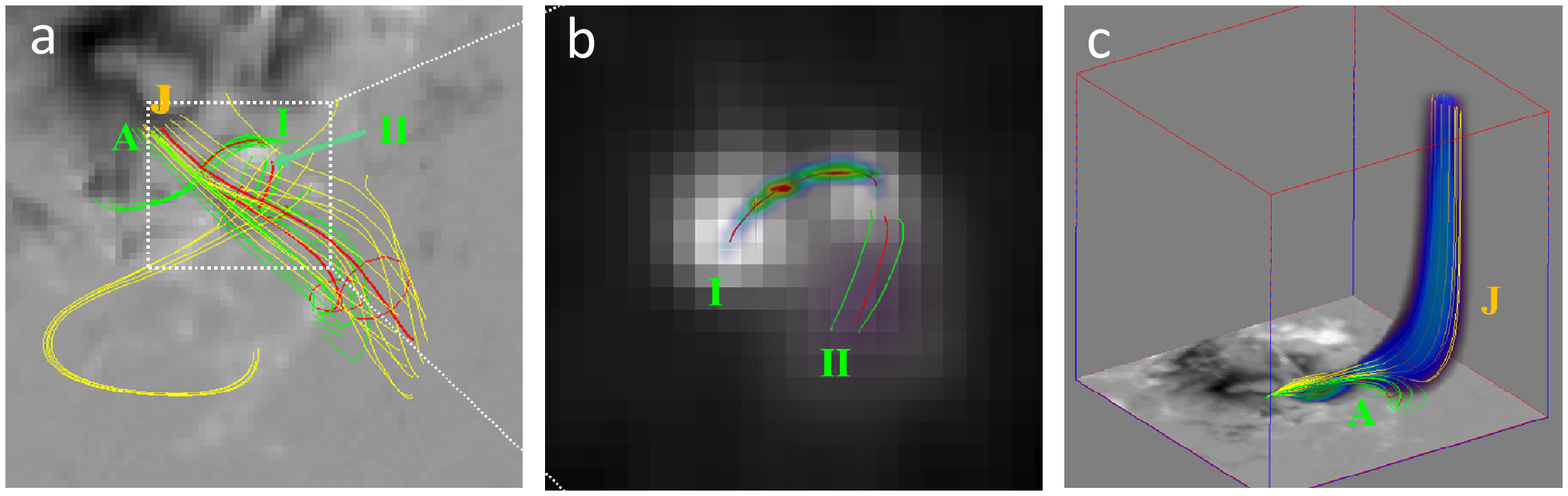}
\includegraphics[width=0.23\textwidth,trim=0 -1mm 0 0, clip]{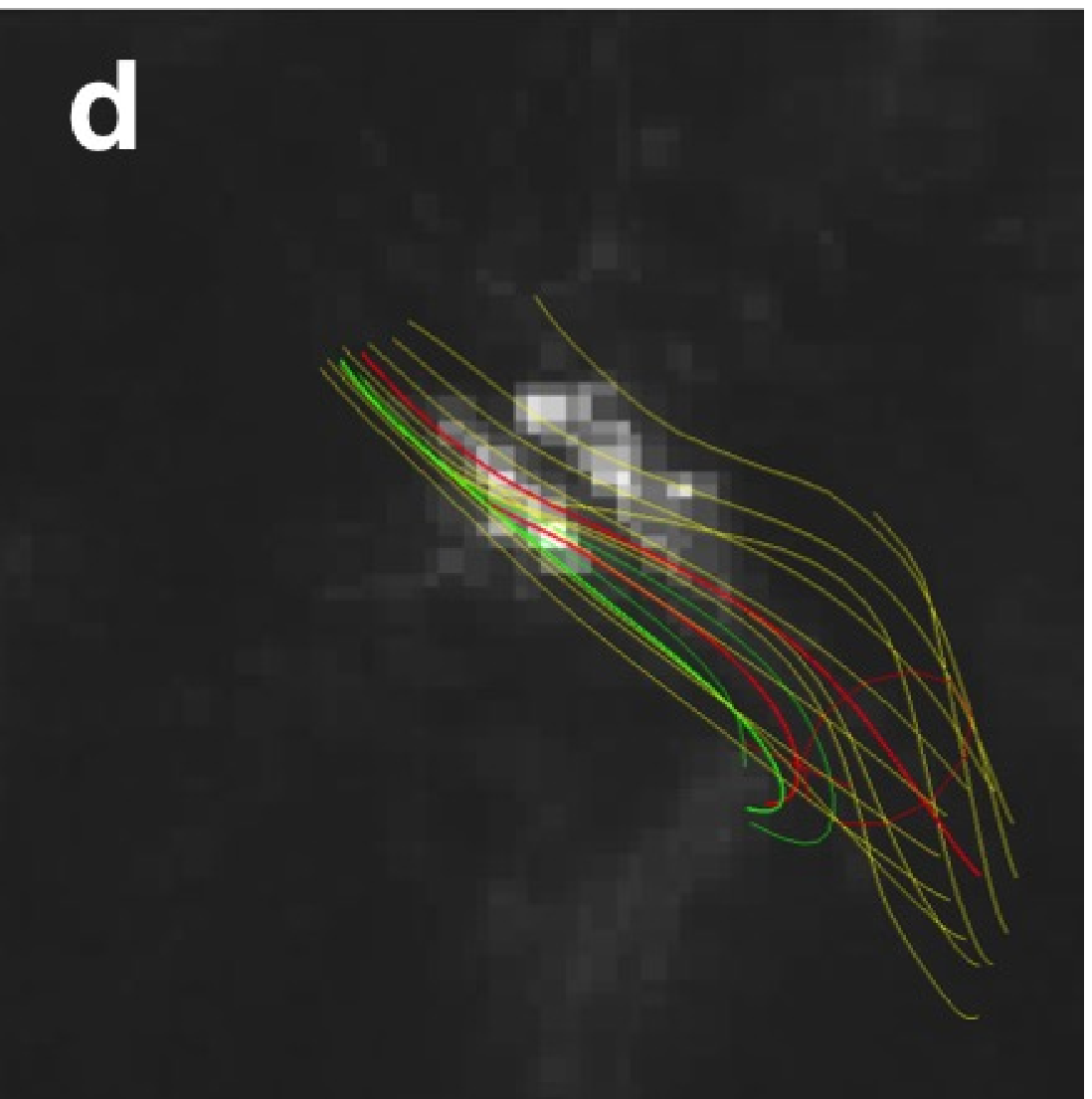}
\caption{\label{f_perspective_jet_model}   Modeled magnetic field geometry and spatial distribution of nonthermal electrons in GX Simulator.  LFFF extrapolations were performed on MDI data in two sets with different values of $\alpha$, including closed,  twisted field connecting flare footpoints (loops marked I and II), and the open, low-twist field along which the jet emerges (marked J), as well as additional closed field (marked A).  Red lines trace a few field lines to serve as ``center'' (reference) lines of the flux tubes.   Yellow lines indicate open field, while green lines indicate closed.  Panel (a) shows a composite view of all relevant flux tubes over the MDI data from which they were extrapolated.  In Panel (b), a zoomed-in view shows the nonthermal electron distribution (green volume) on the closed loops I and II overplotted on a \rhessi\ 25-80~keV image.   In Panel (c), a different perspective view shows more clearly the escape path of the jet J.  The modeled accelerated electron distributions at the jet and adjacent closed loop are illustrated with blue and green intensities, providing a visualization of the nonthermal electrons that escape along the jet.   Panel (d) shows the extrapolated field lines (open and closed) superposed on a \trace\ EUV image.
}
\end{figure*}

\begin{figure*} \centering
\includegraphics[width=0.75\textwidth,clip]{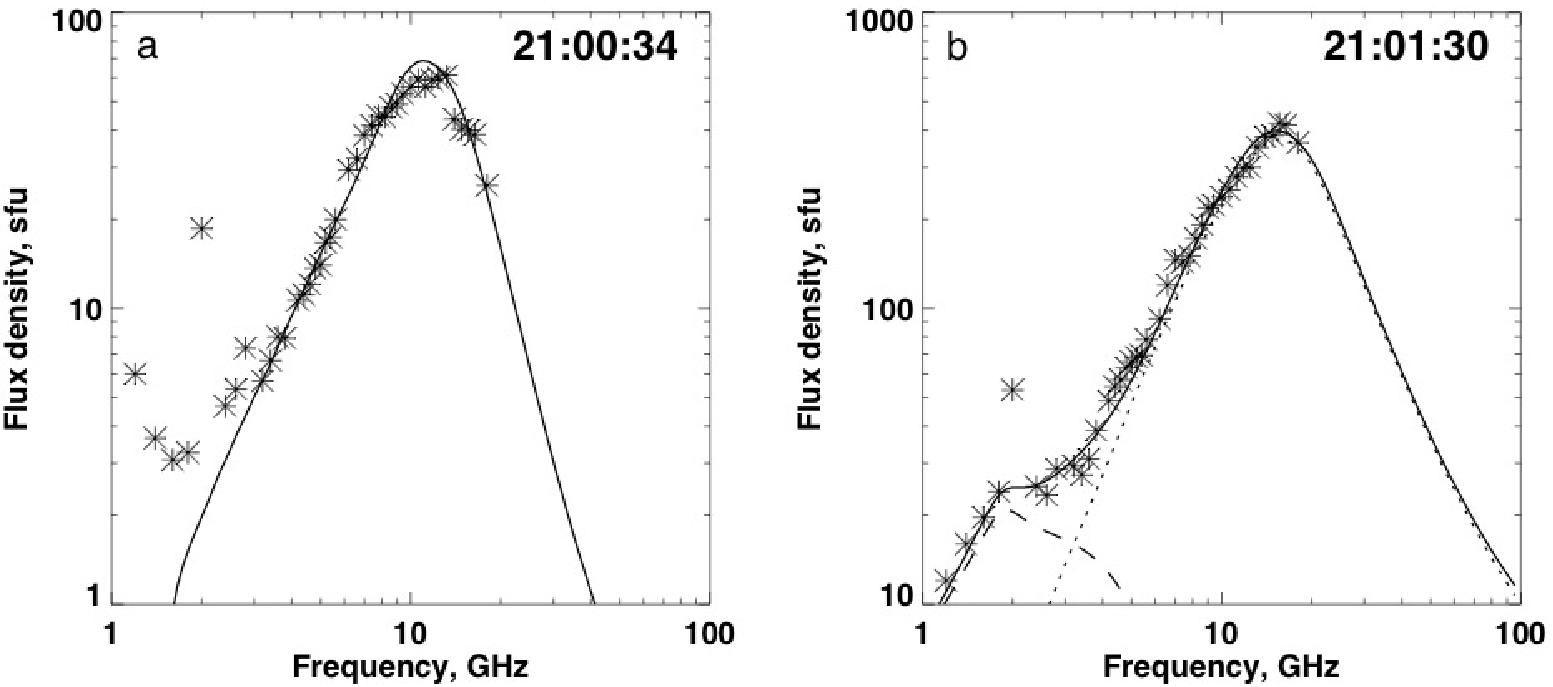}
\includegraphics[width=0.75\textwidth,clip]{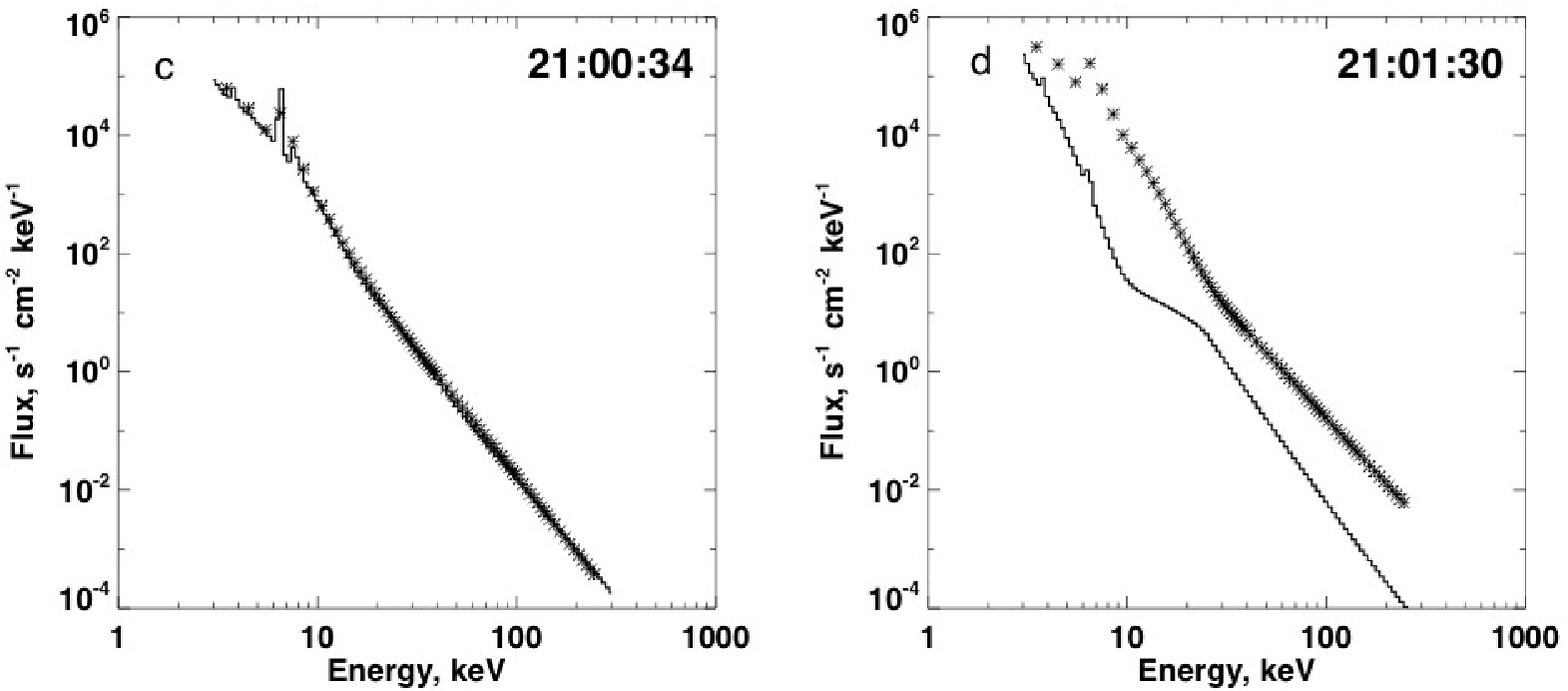}
\caption{(Top row) Observed microwave total power spectra at two time frames: 21:00:34~UT (left) and 21:01:30~UT (right) and the corresponding synthetic spectra obtained from the 3D models using GX Simulator.  The dotted line in the upper right panel shows the contribution computed from the closed flux tube, while the dashed line shows the contribution from the jet itself, necessary to reconcile the observed data at low frequencies.  (Bottom row) Observed and modeled X-ray emission for the two time intervals.
In the bottom right, the model shown by the solid line considers only the contribution from the jet-related part of the model, Fig.~\ref{f_perspective_jet_model}(c); its contribution to the total X-ray spectrum is small throughout the spectrum.  The relative contribution of the jet electrons to the predicted X-ray spectra agrees with \rhessi\ imaging data, which reveal the jet only (and barely) in the 18--30 keV range.
\label{f_20020819_model_spec}
}
\end{figure*}

\subsection{Approach to magnetic field modeling}
\label{sec:mdi}

The 195~\AA\ \trace\, EUV images  illustrate  a complex magnetic topology, which is unlikely to be recovered using either a potential field (PF) or a linear force-free field (LFFF) model of the coronal magnetic field; thus, a nonlinear force-free field (NLFFF) model would be the most suitable for 3D magnetic field modeling, as in \citet{2018ApJ...852...32K}. However, NLFFF modeling requires vector components of the magnetic field at the the photosphere, which were unavailable at the time of our event.  Instead, we create separate magnetic data cubes for distinct components of the overall magnetic structure. Modeling them individually using LFFF extrapolations allows those different components, which may be closed or open field, to have dissimilar values of the force-free parameter $\alpha$, as in \citet{Fl_etal_2016coldFl}.
In particular, different models must be developed for the (presumably highly variable) \mw\ emission from the jet and (more gradual) emission from any closed loops, from which most of the soft X-ray emission originates.

Next, we develop two different models for two time frames indicated by dashed vertical lines in Figure~\ref{f_20020819_overview}b.  At time 21:00:34~UT, the \mw\ emission evidences a valley between peaks (presumably a slowly varying pedestal due to a trapped component), while at 21:01:30~UT (close to the peak of the burst) the jet produces a highly variable component.  We refer to these as ``pre-jet'' and ``jet'' intervals.

\subsection{Modeling the pre-jet interval: closed field}

For the pre-jet time (21:00:34~UT) we use the morphology implied by the observed X-ray structure as an initial guess.  We find that the data near the flare site are well represented by an LFFF extrapolation (3D magnetic data cube) with force-free parameter $\alpha=(8\pm2)\times10^{-10}$~cm$^{-1}$.  Our closed field lines computed from the extrapolated 3D magnetic data cube connect all three primary nonthermal HXR sources observed in the event; see Figure~\ref{f_perspective_jet_model}, panels A and B.  
Using one of these field lines as a reference, we
 create a flux tube (flux tube I in Fig.~\ref{f_perspective_jet_model}) connecting two HXR sources, presumably footpoints, and fill this loop with hot, dense, thermal plasma  ($T=29$~MK, $n_0=10^{11}$~cm$^{-3}$) and nonthermal electrons with a looptop number density $n_b=4\times10^{9}$~cm$^{-3}$ above $E_{\min}=25$~keV.  A relatively dense thermal plasma is needed to avoid, via Razin suppression, otherwise strong \mw\ emission at frequencies below 10~GHz.  The total number of nonthermal electrons integrated over the loop volume is $\int n_bdV\approx 2.8\times10^{34}$.   Any larger  nonthermal number density would overestimate the HXR flux compared with that observed, while a larger thermal density or lower temperature would overestimate the \trace\,  195~\AA\ emission\footnote{A routine computing the EUV \trace\ emission has been recently added to GX Simulator SSW distribution.}.  Taking the nonthermal electron spectral index to be $\delta = 4.8$, which ensures the correct high-frequency slope of the microwave emission, we find that the modeled microwave spectrum matches the observed one; see Fig.~\ref{f_20020819_model_spec} (top left). This tells us that the nonthermal electrons trapped in the loop connecting the HXR footpoints do play a role in forming the slowly-varying microwave emission, which may form the pedestal on top of which the highly variable emission is superimposed. This flux tube alone is insufficient to reproduce the thermal X-ray emission. Thus, we added one more thermal loop with number density $n_0=6\times10^{10}$~cm$^{-3}$ ($EM$ $=5.16\times10^{47}$~cm$^{-3})$ and temperature $T$ $=25$~MK, consistent with those obtained from the \rhessi\ fit. The components together well reproduce both the low-energy imaging data and the spectrum (Fig.~\ref{f_20020819_model_spec}, bottom left).  The agreement between the model and data is perfect due to the recently added ability of GX Simulator to for account collisions not only with hydrogen but with other atoms, as well as free-free and free-bound transitions, various line emissions in the thermal part of the spectrum, and thick-target emission in the nonthermal part of the spectrum.
The data cube so far, however, does  \textit{not} include  an open magnetic flux tube matching the EUV jet.

\subsection{Modeling the jet: open flux tube}

To create the jet-like open flux tube requires a less twisted magnetic structure, which we illustrate by creating a separate model for the second time frame, 21:01:30~UT. To produce the required magnetic data cube we employ an LFFF extrapolation with force-free parameter $\alpha=(1.5\pm0.8)\times10^{-10}$~cm$^{-1}$ (almost PF); see Figure~\ref{f_perspective_jet_model}, panel C. We populate this open flux tube with moderately dense hot thermal plasma ($T=2.05$~MK, $n_0=10^{10}$~cm$^{-3}$; needed to make the jet visible in the 195~\AA\ channel) and nonthermal electrons ($n_b=1.5\times10^{7}$~cm$^{-3}$ above $E_{\min}=25$~keV, and spectral index $\delta = 5$).  The total number of nonthermal electrons integrated over the jet volume is $\int n_bdV\approx 2.9\times10^{34}$.  The spectral index of the nonthermal electrons at the jet is not well constrained by the data but its exact value has only a minor effect on the optically thick low-frequency emission. The nonthermal number density and spatial extent of the nonthermal electrons are reliably constrained by the data at least within a factor of a few.  Microwave emission computed from this jet model yields the correct location and also offers a good match to the low-frequency (optically thick) portion of the microwave spectrum (dashed curve in Fig.~\ref{f_20020819_model_spec}, top right), but cannot account for the high-frequency part of the spectrum, because the magnetic field is too small along the jet.   To account for this high-frequency component  we created a closed flux tube, adjacent to the open flux tube, but located lower in the corona (green field lines in Figure~\ref{f_perspective_jet_model}, panels C and D).
We added a fraction of nonthermal electrons to this closed flux tube where the magnetic field is reasonably strong: $n_b=3\times10^{8}$~cm$^{-3}$ above $E_{\min}=25$~keV, and spectral index $\delta = 4.2$.  The total number of nonthermal electrons integrated over the loop volume is $\int n_b dV\approx 1.4\times10^{34}$.  This offers a good match to the high-frequency part of the microwave spectrum; see the dotted curve in Panel B of Figure~\ref{f_20020819_model_spec}.
  In all cases we computed the EUV emission in the 195~\AA\ passband and made sure that the model does not overestimate the observed EUV emission.

Although it is not possible to perform detailed spectroscopy on the faint HXRs from the jet itself given the bright thermal and nonthermal HXR emission from the flare site, the integrated \rhessi\, emission provides upper limits on the jet HXRs.  From the images in Figure \ref{f_Jet_2002_08_19_images}, the low energy ($\lesssim$18 keV) and high energy ($\gtrsim$30 keV) emission probably emanates from thermal flare plasma  and chromospheric footpoints, respectively; these are the usual features observed in HXR flares.  Only the ``mid'-range'' emission (18-30 keV) shows a hint of HXR extension along the jet.  Our modeled electron distribution for the jet is consistent with these observations.  The black histogram in the lower-left panel of Figure \ref{f_20020819_model_spec} represents the synthetic jet HXR emission and lies $\sim$2 orders of magnitude below the synthetic emission from the closed loops across most of the energies except for a knee around the electron low-energy cutoff at 25 keV.  Near that energy, the synthetic jet emission comes within an order of magnitude of the closed-loop emission, making it barely imageable within \rhessi's imaging dynamic range.  Note that the low-energy cutoff is not well constrained, but the \rhessi\, emission gives us confidence that it lies between 18 and 30 keV as it would be difficult to find another way for the jet HXRs to be barely imageable by \rhessi\, only in that range.  (A ``knee'' is necessary.)

\subsection{Observational constraints and the allowed parameter ranges for the modeled distributions}

Table 1 gives a summary of the GX Simulator electron distributions including thermal and nonthermal parameters for all of the modeled loops, both at the initial time (21:00:34 UT) when only two twisted loops are included, and at the later time of the jet (21:01:30 UT) when the structure has changed to an open set of field lines (along which the jet emerges) and one closed loop.  Here, we discuss the parameter ranges and how they are constrained by the broad set of observations.  We do not guarantee that this model is a unique solution; it is conceivable that other flux tube geometries and other spectral assumptions (besides our assumed gyrosynchrotron model) could fit the data.  However, after an extensive (though non-exhaustive) effort to seek believable alternatives, we did not find other configurations or models that well represented all the data sets.

The magnetic field lines were extrapolated from \textit{SOHO}/MDI data using LFFF extrapolation, with force-free parameters selected so as to reproduce the magnetic connectivity suggested by the data, as explained in Section \ref{sec:mdi}.  The parameters given are the best-fit values based on these extrapolations.  The force-free parameter has an overall ``nominal'' value and varies slightly along each flux tube due to numerical effects in the LFFF modeling; this range of variance is given in the table.  In all cases, the extrapolated field lines connect the HXR footpoints and match the location of heated plasma observed by \trace. \\

\subsubsection{Time interval 1: Loops I and II.}

For Loops I and II we use thermal parameters (emission measure $EM$, temperature $T$) constrained by spectral fits to \rhessi\, integrated data, so the thermal plasma parameters closely agree with the X-ray data. Nonthermal parameters in the tens-of-keV range (cutoff energy $E_0$ and spectral index $\delta_1$) are also fit from an integrated \rhessi\, spectrum.  As is often the case in fitting \rhessi\, integrated spectra, the low-energy cutoff of the electrons is poorly constrained in the presence of bright thermal plasma.  The fit value (25 keV) is best viewed as an upper limit on this parameter.  To ensure that our results do not hinge on our adjusting this parameter, we use 25 keV as the low-energy cutoff for \textit{all} nonthermal electron distributions considered in this paper.  If the value is truly lower then even more energy is present in accelerated electrons.  (See Section 3 of \citet{holman2011} for a thorough discussion.)
The loop width is taken to match the low-frequency, optically thick part of the \mw\ emission.  This emission is fully defined by the product of only two parameters---source area and brightness temperature. The latter is fully defined by the magnetic field (fixed for a given flux tube) and parameters of the nonthermal electron distribution (fixed from the X-ray fit); meaning that the loop area can be derived from the \mw\ data.

The remaining parameters are: the break energy, high-energy spectral index, and the maximum energy; all of these high-energy parameters are constrained by the \mw\ data.  The observed \mw\ spectrum is well reproduced with a single power-law electron distribution ($\delta=4.8$, see Table~\ref{Table_parms_full}) if the maximum electron energy $E_{\max}\sim400$~keV.  If we instead permit the maximum electron energy to be larger than $E_{\max}>500$~keV, then a double power-law model is needed with a break around 300~keV.  In other words, a softening of the distribution above 300 keV is necessary, whether by a break in the power law or a hard cutoff to the distribution.  The details of this softening have a negligible effect on our computation of total electron energy or in the interpretation of the scenario, but a single power-law extending to the energies above 500~keV is excluded.  Table \ref{Table_parms_full} lists parameters for only a single power law. \\

\subsubsection{Time interval 2: Jet plus adjacent closed loop.}

The X-ray, EUV, and microwave data work together to constrain the thermal distribution at this time.  The jet density must be below $\sim1.7\times10^{10}$~cm$^{-3}$ in order that the cutoff frequency is below $f=1.2$~GHz (where the \mw\ spectrum starts).  We approximate this density in our model as $n_0=1\times10^{10}$~cm$^{-3}$.  This density could be adjusted, but should not be orders of magnitude lower or the jet would not be observed in the EUV given the \trace\ 195\AA\ response and reasonable temperatures.

For the closed loop, the thermal density is taken to be $n_0=2\times10^{11}$~cm$^{-3}$.  Values larger than this would violate three observational constraints:  (1) 195~\AA\ emission would be stronger than observed for all reasonable coronal (including flaring) temperatures; (2) SXR emission would be stronger than is observed; and (3) the \mw\ spectrum would be dominated by a free-free component, which would not match the observed \mw\ spectral shape. The density could be smaller but would require adjustment of the loop geometry; therefore, there is a range of thermal densities that could fit the observed data.  For the likely value of $1\times10^{10}$~cm$^{-3}$ for the jet density, a temperature of 2.05 MK fits the measured \trace\ 195~\AA\ emission from the jet location.  Since the \trace\ 195~\AA\ temperature response function \citep{handy1999} decreases for all higher temperatures, a higher density would be required for any jet temperatures higher than this (and would violate the density constraint from \mw\ data mentioned earlier).

For the closed loop, a density of $n_0=2\times10^{11}$~cm$^{-3}$ can produce the observed 195~\AA\ brightness at the jet base, where this loop is located, for a temperature between 7.3 and 10.6~MK.
Since the plasma temperature is not constrained within this range, we choose a temperature of 7.3 MK.

For the nonthermal electrons in the jet, the \mw\ spectrum deviates from observations if the upper cutoff to the electron energy $E_{\max}$ is below $\sim300$~keV; it is not well constrained from above.  Further decrease of this parameter could be compensated by an increase of the source area, but this would be in conflict with the \mw\ source size.  For the closed loop the \mw\ spectrum deviates from observations if $E_{\max}< 0.8$~MeV. In the model we use $E_{\max}=0.9$~MeV.

\begin{deluxetable*}{|ll|cc|cc|}
\tablecolumns{5}
\tablewidth{0pc}
\tabletypesize{\footnotesize}
\tablecaption{Field and electron distributions modeled in GX Simulator based on MDI, \rhessi, OVSA, and \trace\, data.  }
\tablehead{\colhead{Parameter}& \colhead{Symbol, units} & \colhead{Loop I} & \colhead{Loop II} &  \colhead{Open jet } & \colhead{Adjacent loop} }
\startdata
 \multicolumn{2}{c|}{Time} & \multicolumn{2}{c|}{21:00:34 UT} & \multicolumn{2}{c}{21:01:30 UT} \\
\hline
{\textit{Central field line}:} &  & & & & \\
\quad Length      & $l$, cm  & $1.875\cdot10^9$ & $1.27\cdot10^9$ & $1.83\cdot10^{10}$ & $5.0\cdot10^9$ \\
\quad Force-free parameter: Nominal &
                                        $\alpha/ (10^{-10}$cm$^{-1}$) &  $8.18$ &  $8.18$ & 1.36 & 1.36\\
\quad \quad \quad \quad Along the flux tube
                                            & $\alpha/ (10^{-10}$cm$^{-1}$) &  $8\pm2$ &  $8\pm2$ & $0$  & $1.5\pm0.8$ \\
\quad Number of twists & $N_{twist}={\alpha l}/({4\pi})$ &  $0.12$ &  $0.08$ & $0$ & $0.06$\\
\quad Magnetic field, positive footpoint      & $B_{f+}$,~G  & 655 & 292 & 258 & 423 \\
\quad Magnetic field, negative footpoint      & $B_{f-}$,~G  & -397 & -585 & -3625 & -768  \\
\quad Magnetic field, looptop      & $|B_{\rm ref}|$,~G  & 294 & 268 & 102 & 111  \\
\textit{Flux tube:} &  & & & & \\
\quad Reference cross-section radius\tablenotemark{a} & $a=b$, cm & $1\cdot10^8$ & $5.87\cdot10^8$ & $8.0\cdot10^8$  & $6.2\cdot10^8$\\
\quad Model volume; $\left[\int n_0 dV\right]^2/\int n_0^2 dV$ & $V$, cm$^3$ & $2.43\cdot10^{25}$ & $5.64\cdot10^{26}$ & $8.35\cdot10^{27}$ & $1.77\cdot10^{27}$ \\
{\textit{Thermal plasma}:} &  &  &  & &  \\
\quad Number density at central field line & $n_0$,  cm$^{-3}$ & $1.0\times10^{11}$ & $0.6\times10^{11}$ & $<1.7\times10^{10}$ & $2\times10^{11}$  \\
\quad Emission Measure; $\int n_0^2 dV$ & $EM$,  cm$^{-3}$ &  $3.3\times10^{46}$ & $5.16\times10^{47}$ & $1.1\times10^{47}$& $4.46\times10^{48}$  \\
\quad Mean number density; $\int n_0^2 dV/\int n_0 dV$ & $\langle n_0\rangle$,  cm$^{-3}$ & $3.7\times10^{10}$ & $3.0\times10^{10}$ & $3.62\times10^{9}$ & $5.0\times10^{10}$  \\
\quad Temperature      & $T$, MK  &  29 & 25  &  $>2.05$ & 7.3--10.6  \\
\quad Parameters of transverse distribution\tablenotemark{*}      & $p_0,~p_1,~p_2,~p_3$  &  2, 2, 0, 0 & 2, 2, 0, 0   &  2, 2, 0, 0 & 0.5, 0.5, 0, 0 \\
\quad Parameters of distribution along the loop\tablenotemark{**}      & $q_0,~q_1,~q_2$  &  barometric\tablenotemark{$\dag$} & barometric\tablenotemark{$\dag$}  &  barometric\tablenotemark{$\dag$}  &  5, 0, -0.9   \\
\textit{Nonthermal electrons:} &  &  &  & &   \\
\quad Number density at central field line & $n_b$, cm$^{-3}$ &  $4.0\times10^{9}$ &  --- &  $1.5\times10^{7}$  &  $3.0\times10^{8}$   \\
\quad Mean number density; $\int n_b^2 dV/\int n_b dV$ & $\langle n_b\rangle$,  cm$^{-3}$ & $1.37\times10^{9}$ &  --- & $5.08\times10^{6}$  & $1.06\times10^{8}$  \\
\quad Parameters of transverse distribution\tablenotemark{$\ddag$}      & $p_0,~p_1,~p_2,~p_3$  &  2, 2, 0, 0 &  ---  &  2, 2, 0, 0  &  2, 2, 0, 0 \\
\quad Parameters of distribution along the loop\tablenotemark{$\flat$}      & $q_0,~q_1,~q_2$  &  1, 2.9, -0.3 &  ---  &  5, 0, 0.2  &  11, 0, -0.7   \\
\quad Total electron number & $N_b$, cm$^{-3}$ &  $2.8\times10^{34}$ &  ---  &  $2.9\times10^{34}$ &  $1.4\times10^{34}$   \\
\quad Low-energy cutoff & $E_0$, keV & 25  &  ---  & 25 & 25  \\
\quad Maximum electron energy  & $E_{\rm max}$, MeV & 0.4 &  ---  & $>0.3$ & $>0.8$ \\
\quad Electron spectral index & $\delta_{n1}(<E_{\rm break})$ & 4.8 &  ---  & 5 & 4.2 \\
\enddata
\tablenotetext{*}{In our case, the reference location is chosen to be that of minimum magnetic field value.}
\tablenotetext{*}{The distribution is described by Equation~(\ref{Eq_n0_xy}).}
\tablenotetext{**}{\ The distribution is described by Equation~(\ref{Eq_n0_s}).}
\tablenotetext{$\dag$}{The distribution is described by Equation~(\ref{Eq_n0_z}).}
\tablenotetext{$\ddag$}{The distribution is described by Equation~(\ref{Eq_nb_xy}).}
\tablenotetext{\flat}{The distribution is described by Equation~(\ref{Eq_nb_s}).}
\label{Table_parms_full}
\end{deluxetable*}

\section{Discussion}

To recap the observations, we find high time variability in HXRs and microwaves, with HXR spectral hardening strongly correlated with fast bursts of intense flux.  HXR images are dominated by the flaring (thermal) loops at low energies and (nonthermal) footpoints at high energies, but at intermediate energies ($\sim$18--30 keV), some \rhessi\, emission is elongated along the jet, where OVSA data also reveal an elongated microwave source at the location of the jet.

We interpret the fast HXR spikes as bursts of particle acceleration, and the duration of these peaks gives the timescale over which electrons are accelerated.  The very short time lag between the burst intensity and its hardening (less than 200~ms in the case of G1 and no lag in the case of G2) implies that electrons are accelerated quickly to $\sim100$~keV.  Combining the time lag from Figure \ref{f_cross_correlation} with the peak width from Figure \ref{fig:autocorrelation}, we conclude that the event exhibits electron energy increases on an average timescale of 0.2 seconds, with acceleration episodes lasting, on average, 1 second.  Once accelerated, the electrons that have access to open field simply escape, and so there is no lengthening of HXR pulse durations due to magnetic trapping.  We surmise that the short pulses are associated with the acceleration of electrons into the jet.  Since these electrons escape, many injections over a substantial time interval are needed to replenish the electrons and produce observable emission; this is consistent with the high variability lasting for $\gtrsim$1 minute in the HXR and \mw\ time profiles.  For accelerated electrons that are injected into the flare loop, on the other hand, trapping could lengthen the duration of  associated HXR  and radio components; this could account for the $>10$ second timescales in the \kw\, data  and the pedestal in the \mw\ emission.  We note that \citet{kiplinger1983} performed a search for fast HXR variations, finding them in $\sim$10\% of the flares studied.  \citet{qiu2012} and \citet{cheng2012} also found $<$1 second HXR peaks in demodulated \rhessi\, data.  It is difficult to tell from these past studies (using data taken before the \textit{SDO} era, and only occasionally overlapping with \trace\ coverage) whether these quickly-varying events were systematically associated with jets.  Future work will explore this subject using more recent observations.

The variability observed in OVSA data is unusual for a microwave burst but when observed is interpreted as pulsed or beam-like acceleration of electrons \citep[e.g.][]{Altyntsev_etal_2008, Fl_etal_2008}.  This behavior is difficult to reconcile with variability of  a single nonthermal electron population due to transport effects in a given magnetic flux tube, and instead either requires a number of distinct sources (loops) or a sequence of distinct acceleration / injection episodes in a single source.
However, strong variability of the spectral peak frequency, sensitive to the magnetic field value at the source, favors the scenario with multiple sources---loops with accordingly different magnetic field magnitudes.  Most of the \mw\ emission comes from the closed loop (A), while a contribution from the open field with a lower magnetic field magnitude is needed to account for the low-frequency spectral knee.  While gyrosynchtron is not the only mechanism that can produce broadband \mw\ emission, we do not find reasonable alternatives for this event.  We attempted fits to the \mw\, data using a thermal bremsstrahlung model and found that this scenario would require unphysically large densities or source sizes; in either case, the emission would violate \rhessi\ observational constraints.  Additionally, cooling and heating timescales in the corona would not allow the observed fast ($\lesssim$ 1 second) time variations for a thermal population.

The question of symmetry (or lack thereof) in numbers of flare electrons accelerated upward (toward interplanetary space) versus downward (toward the chromosphere) is important for flare acceleration theories, but is poorly understood to date.  Some studies of in-situ electrons at 1 AU find that escaping electrons represent only a minor ($<1\%$) fraction of the electrons accelerated in flares \citep{lin1971, krucker2007}.  However, more recent work found similar energies in electrons accelerated upwards and downwards \citep{james2017}.  In our event, combination of the data and 3D modeling allows us to estimate the nonthermal electron population in the open, jet-forming flux tube, finding ($\sim 3\times10^{34}$) electrons (above 25 keV) on the open field.  This is comparable to the number of electrons in the closed flux tube.  Our result is in line with the work of \citet{james2017} and also with a recent finding of \citet{Fl_etal_2016coldFl}, who analyzed a flare  produced by an interaction between two loops---one small and one large. In  that flare the accelerated electrons were divided roughly equally between the two loops.

\section{Conclusions}

In summary, we have used a combination of HXR, EUV, and radio data, combined with modeling of the emission from  thermal and accelerated
electron populations, to form a credible spatial and energetic distribution for the accelerated electrons in a solar
jet. The direct microwave detection and the HXR upper limits are needed to construct the simulated accelerated electron distribution. As far as
we are aware, this is the first case in which microwave gyrosynchrotron emission has been detected from an open, rather
than closed, magnetic configuration, and is the most direct constraint to date on the accelerated electron population within a
solar jet. It is extremely important that the model built using actual magnetic field data yields an excellent match of the simulated and
observed radio image and spectrum, thus validating and quantifying the nonthermal electron distribution on the open field flux tube. We
stress that for the identification and analysis of such an event, the necessary approach was the careful consideration of HXR, EUV,
and radio data combined with modeling.

Jets occupy an important role in solar and heliospheric physics, as they provide a direct way for impulsive solar events to influence the heliosphere.  We expect the study of jets to become even more prominent with the promise of in-situ measurements by \textit{Solar Probe Plus} and \textit{Solar Orbiter}, which will measure the energetic particles that reach the heliosphere directly and through their Type III radio emission.  Future investigation using cutting-edge instruments will utilize direct imaging of solar HXRs -- for example using the technology from the successful \textit{FOXSI} rocket program \citep{krucker2014, glesener2016} -- as well as the microwave imaging spectroscopy offered by EOVSA \citep[e.g.][]{wang2015}.  With these instruments, the escaping electrons could be imaged at their source and these measurements could be compared with those at multiple points in the heliosphere, allowing for a complete picture of electron acceleration and transport.

\acknowledgments

This work was supported in part by NSF grants
AST-1615807 and AGS-1817277,  NASA grants
NNX16AL67G, 80NSSC18K0015,
80NSSC18K0667  to the New Jersey
Institute of Technology, and by an NSF Faculty Development Grant (AGS-1429512) to the University of Minnesota.  The authors are grateful to Alexandra Lysenko and the \kw\ team for making their data available and to Dale Gary, Gelu Nita, S\"{a}m Krucker, and Sophie Musset for insightful comments on the text.

\bibliographystyle{apj}
\bibliography{fleishman,fleishman_old,jet_ref}

\appendix
\section{Some definitions in GX Simulator}
\label{S_GX_def}

\subsection{Thermal plasma distribution.}

The distribution of the thermal electrons along the spine of the flux tube (i.e., along the reference field line shown in red in Figure \ref{f_perspective_jet_model}a) may be a hydrostatic equilibrium (barometric) distribution, which is the default, in which case it is controlled by temperature $T_0$:

\begin{equation}
\label{Eq_n0_z}
n_{0}(s(z))=n_0\exp\Big\{-\frac{z}{6.7576\times10^{-8}R_{\sun} T_0}\Big\},
\end{equation}
where $n_0$ is the maximum number density along the field line (in this case, at the footpoint), $z$ is the height above the photosphere, and $R_{\sun}$ is the solar radius.  Alternatively, $n$ can be an arbitrary, user-defined, function of $s$, the position along the loop. If the default distribution is inconsistent with the data, we employ the following, generalized Gaussian distribution, which we also use in the case of nonthermal electrons (see below):

\begin{equation}
\label{Eq_n0_s}
n_{0}(s)=n_0\exp\Big\{-\Big[q_0\Big(\frac{s-s_0}{l}+q_2\Big)\Big]^2-\Big[q_1\Big(\frac{s-s_0}{l}+q_2\Big)\Big]^4\Big\},
\end{equation}
where $l$ represents the length of the loop central field line,  the parameter $q_0$ controls the width of the Gaussian, and $q_2$ indicates where the gaussian has a maximum relative to a pre-selected $s_0$ reference point (e.g., the loop-top)\footnote{The adopted convention is that the coordinate $s$ along the central (reference) field line of the flaring flux tube is zero where the magnetic field is at its minimum in absolute value.  $s=s_{\min}<0$ at the footpoint of the positive magnetic field, while $s=s_{\max}>0$ at the footpoint of the negative magnetic field.  The difference $s_{\max}-s_{\min} = l$, where $l$ is the field line length.}.

For each location $s$ along the fluxtube spine, the transverse distribution is defined as a two-dimensional (2D) generalized Gaussian function
\begin{equation}
\label{Eq_n0_xy}
n_{0}(x,y,s)=n_0(s)\exp\Big\{-\Big(\frac{p_0 x}{a(s)}\Big)^2-\Big(\frac{p_0 y}{b(s)}\Big)^2-\Big(\frac{p_0 x}{a(s)}\Big)^4-\Big(\frac{p_0 y}{b(s)}\Big)^4 \Big\},
\end{equation}
where $a(s_0)$ and $b(s_0)$ represent semi-axes of the reference elliptical cross-section of the flux tube located at $s=s_0$; $a$ and $b$ scale together with $s$ following magnetic flux conservation such as $B(s)a(s)b(s)=$const.

\subsection{Nonthermal electron distribution:}

The default distributions of the nonthermal electrons along and transverse to the flux tube spine are the same as Eqns.~\ref{Eq_n0_s} and \ref{Eq_n0_xy} for the thermal electrons:

\begin{equation}
\label{Eq_nb_s}
n_{b}(s)=n_b\exp\Big\{-\Big[q_0\Big(\frac{s-s_0}{l}+q_2\Big)\Big]^2-\Big[q_1\Big(\frac{s-s_0}{l}+q_2\Big)\Big]^4\Big\},
\end{equation}

\begin{equation}
\label{Eq_nb_xy}
n_{b}(x,y,s)=n_b(s)\exp\Big\{-\Big(\frac{p_0 x}{a(s)}\Big)^2-\Big(\frac{p_0 y}{b(s)}\Big)^2-\Big(\frac{p_0 x}{a(s)}\Big)^4-\Big(\frac{p_0 y}{b(s)}\Big)^4 \Big\}.
\end{equation}
An important point is that the parameters $q_i$ and $p_i$ are defined independently for the thermal and nonthermal electrons; thus, they do not necessarily coincide for the given flux tube.

\end{document}